\documentclass[prl,twocolumn,showpacs,preprintnumbers,amsmath,amssymb,superscriptaddress]{revtex4}

\usepackage{graphicx}
\usepackage{dcolumn}
\usepackage{bm}

\begin{document}

\title{Compression of sub-relativistic space-charge-dominated electron bunches for single-shot femtosecond electron diffraction}

\author{T. van Oudheusden}
\author{P.\,L.\,E.\,M. Pasmans}
\author{S.\,B. van der Geer}
\author{M.\,J. de Loos}
\author{M.\,J. van der Wiel}
\author{ O.\,J. Luiten}\email{O.J.Luiten@tue.nl}

\affiliation{Department of Applied Physics, Eindhoven University of Technology, P.O Box 513, 5600 MB Eindhoven, The
Netherlands}

\date{\today}

\begin{abstract}
We demonstrate compression of $95\,\rm{keV}$, space-charge-dominated electron bunches to sub-$100\,\rm{fs}$ durations. These bunches have sufficient charge ($200\,\rm{fC}$) and are of sufficient quality to capture a diffraction pattern with a single shot, which we demonstrate by a diffraction experiment on a polycrystalline gold foil. Compression is realized by means of velocity bunching as a result of a velocity chirp, induced by the oscillatory longitudinal electric field of a $3\,\rm{GHz}$ radio-frequency cavity. The arrival time jitter is measured to be $80\,\rm{fs}$.
\end{abstract}


\maketitle

\begin{figure}
    \centering
    \includegraphics[width=\columnwidth]{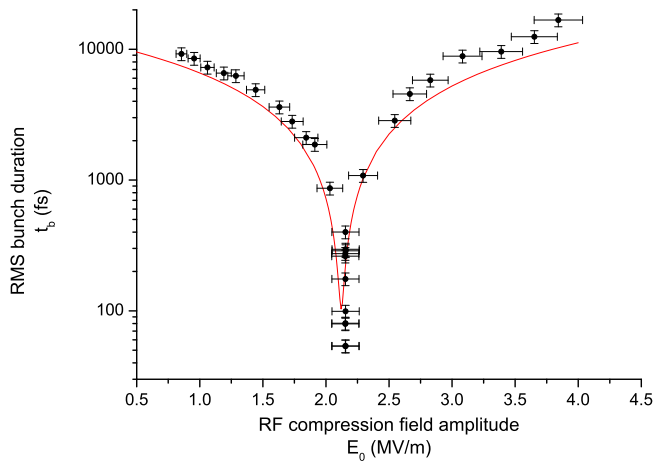}
    \caption{\it (color online) RMS duration $t_b$ of $95.0\,\rm{keV}$, $250\,\rm{fC}$ electron bunches as a function of the RF compression field amplitude $E_0$. GPT simulations (red line) are in close agreement with the measurements (black dots).}
    \label{fig:compression}
\end{figure}

\begin{figure}
    \centering
    \includegraphics[width=\columnwidth]{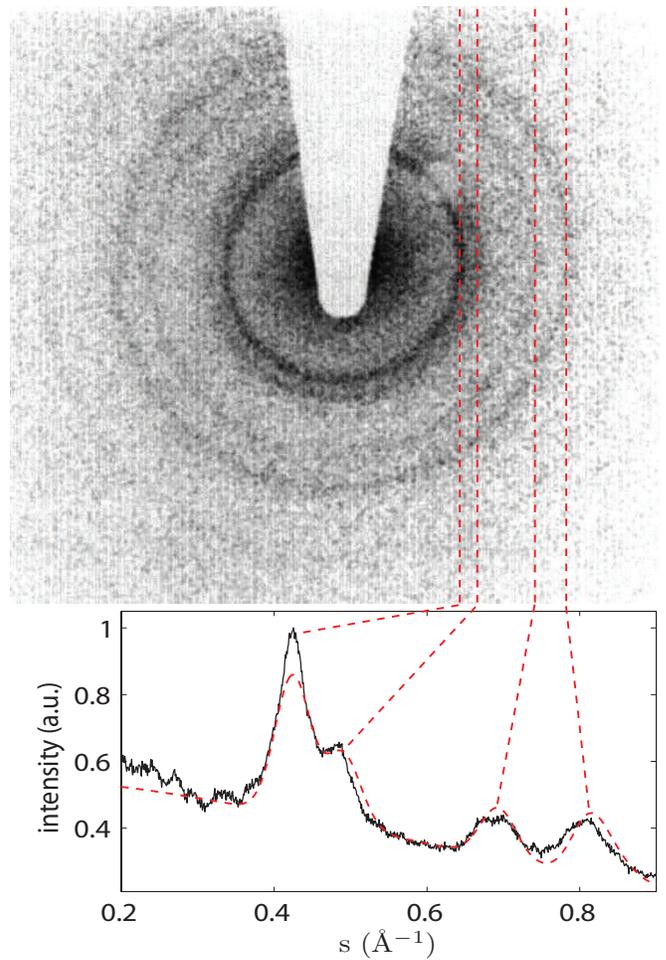}
    \caption{\it (color online) (top) Electron diffraction pattern of a polycrystalline gold foil, recorded with a single $200\,\rm{fC}$ bunch. (bottom) Azimuthal integration of Debye-Scherrer rings (black line) and a fit (red dashed line) based on kinematical diffraction theory.}
    \label{fig:diffraction}
\end{figure}

The breathtaking pace at which ultrafast X-ray and electron science have evolved over the past decade is presently culminating in studies of structural dynamics with both atomic spatial \textit{and} temporal resolution, i.e. sub-nm and sub-$100\,\rm{fs}$ \cite{Dwyer,Zewail-UEDoverview,LCLS-firstlight}. This may revolutionize (bio-)chemistry, and material science and might open up vast new areas of research. In particular, in 2009 the first hard X-ray free electron laser (LCLS) has become operational \cite{LCLS-firstlight}. This has already resulted in single-shot, femtosecond X-ray diffraction experiments on sub-micron crystals of a membrane protein \cite{Chapman-Xraypowder@LCLS}. Over the past few years ultrafast electron diffraction (UED) techniques have been successfully applied to investigate condensed matter phase transitions dynamics at the atomic spatio-temporal scale \cite{Brad,Miller-Bi,Gedik,Ruan}. X-ray diffraction and electron diffraction provide principally different, but in fact complementary, information of atomic structure. Because of their shorter mean free path, however, electrons are favorable for probing thin films, surfaces, or gases. Unfortunately, single-shot, femtosecond operation has not yet been achieved with electrons: because of the repulsive Coulomb force high space charge density bunches will expand rapidly in all directions. To solve this problem the paradigm in the high-brightness electron beam community is to accelerate the electrons to relativistic velocities as quickly as possible. Special relativity dictates that the Coulomb force is then effectively damped, resulting in a slower bunch expansion. Although it has been shown that relativistic bunches can be used for UED \cite{Musumeci-diffraction} they pose difficulties like a reduced cross-section, radiation damage to samples, non-standard detectors, and general expense of technology. However, at the preferred electron energies of $100 - 300\,\rm{keV}$ \cite{Thijs-JAP} the bunch charge required for single shot UED results inevitably in loss of temporal resolution. The obvious solution is to lower the charge per bunch \cite{Brad,Miller-Bi,Zewail-UEDoverview} and use multiple shots to obtain a diffraction pattern of sufficient quality. However, in this way the choice of samples is restricted for reasons of radiation damage and repeatability of the process under investigation. Following this strategy, the closest to single-shot, femtosecond operation has been achieved by Sciaini et al., who used $\sim0.001\,\rm{pC}$ bunches and integrated 4-12 shots per time point to monitor electronically driven atomic motions of Bi \cite{Miller-Bi}. By positioning the sample at $3\,\rm{cm}$ from the photocathode they achieved $350\,\rm{fs}$ resolution. Preferably, however, the bunch charge should be $\gtrsim 0.1\,\rm{pC}$, in particular for UED on more complicated molecular crystals, while maintaining a high beam quality and $\lesssim100\,\rm{fs}$ resolution.
\\
In this Letter we demonstrate 100-fold compression of $0.25\,\rm{pC}$ electron bunches to sub-$100\,\rm{fs}$ durations, see Fig. \ref{fig:compression}. To show that these bunches are of sufficient quality for single-shot UED we have used a single bunch to record the diffraction pattern of a polycrystalline gold foil, as shown Fig. \ref{fig:diffraction}.

The quality of a diffraction pattern is mainly determined by the transverse coherence length $L_c$ of the electron bunch, defined as $L_c = \lambda / \left(2\pi \sigma_{\theta}\right)$, where $\lambda$ is the electron De Broglie wavelength, and $\sigma_{\theta}$ the transverse root-mean-square (RMS) angular spread of the electrons. The transverse coherence length should preferably be larger than the lattice spacing, implying $L_c \gtrsim 1\,\rm{nm}$. Further, the transverse RMS bunch size should preferably be matched to the sample size, which is often limited by sample preparation techniques to $\lesssim 100\,\rm{\mu m}$. When creating the electron bunch by photoemission the combination of the requirements on $L_c$ and spotsize dictates that the RMS radius of the laser at the photocathode should be smaller than $50\,\rm{\mu m}$ \cite{Thijs-JAP}. For a high quality diffraction pattern at least $0.1\,\rm{pC}$ ($\sim10^6$ electrons) is required. When extracting such a charge from a cathode, using a femtosecond laser pulse with the required $50\,\rm{\mu m}$ RMS spotsize, a pancake of electrons is created of which the dynamics are dominated by, generally non-linear, space-charge forces. As a result the bunch will not only expand rapidly, but it will also deteriorate: the RMS angular spread (measured in the beam waist) will increase, leading to a smaller $L_c$. By carefully shaping the transverse intensity profile of the photoemission laser pulse an ellipsoidal bunch can be created \cite{Jom-PRL}, with linear correlations between the velocities and the positions of the electrons \cite{Kellogg}. The expansion of such a bunch is fully reversible with linear charged particle optics, i.e. the transverse coherence length is not affected by space charge forces. To reverse the longitudinal bunch expansion we use radio-frequency (RF) techniques, as we proposed in Ref. \cite{Thijs-JAP}.

\begin{figure*}[tbh]
    \centering
    \includegraphics[width=\textwidth]{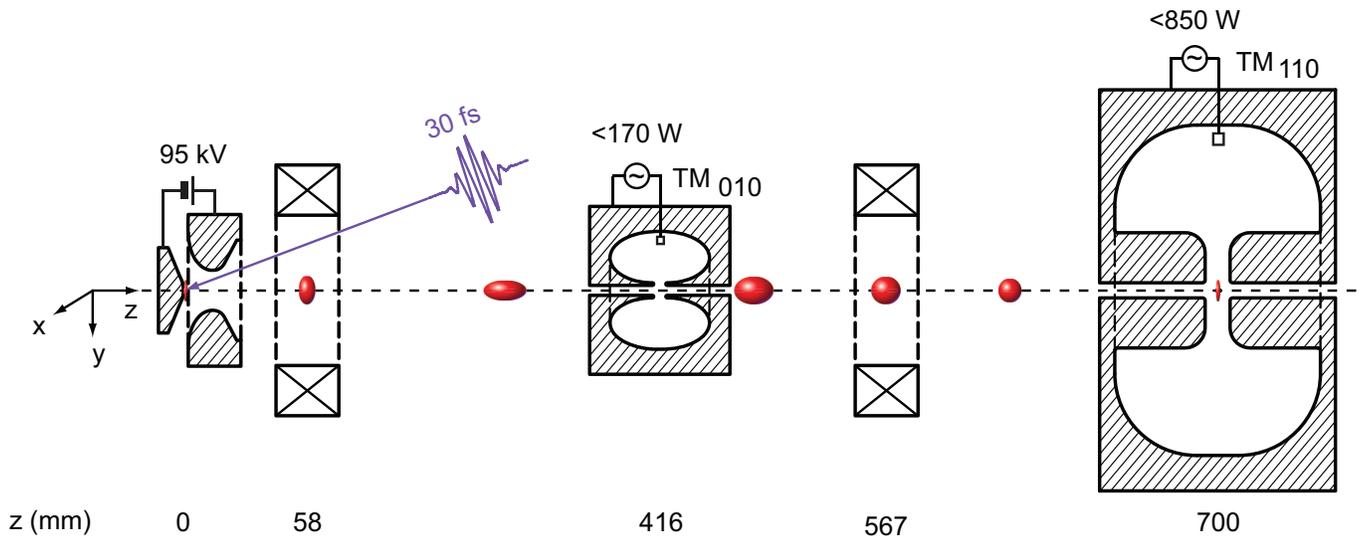}
    \caption{\it (color online) Schematic of the setup and bunch evolution as it propagates through the beamline.}
    \label{fig:setup}
\end{figure*}

Our setup is schematically shown in Fig. \ref{fig:setup} and described in more detail in Ref. \cite{Thijs-JAP}. A (pancake-like) bunch, containing $\geq 10^6$ electrons, is liberated from a copper cathode by photoemission with the third harmonic of a $50\,\rm{fs}$, $800\,\rm{nm}$ Ti:Sapphire laser pulse. The transverse laser profile is Gaussian with a $200\,\rm{\mu m}$ waist, concentrically truncated by a $200\,\rm{\mu m}$ diameter pinhole. The RMS spotsize of the resulting dome-shaped intensity profile is approximately $50\,\rm{\mu m}$. This initial distribution has been demonstrated to evolve into an ellipsoidal bunch \cite{Musumeci-waterbag,Musumeci-longitudinal}.
\\
After photoemission the rapidly expanding electron bunch is accelerated in a DC electric field to an energy of $95.0\,\rm{keV}$. Two solenoids control the transverse bunch size, as illustrated in Fig. \ref{fig:setup}. The first one focusses the bunch through the RF compression cavity and the second one is used to obtain the desired spotsize at the sample. To compress the bunch in the longitudinal direction we use a $3\,\rm{GHz}$ RF cavity oscillating in the $\rm{TM_{010}}$ mode. The RF phase offset is synchronized in such a way that the on-axis longitudinal electrical field $E_z(t)$ inside the cavity decelerates the electrons at the front of the bunch and accelerates the electrons at the back of the bunch, leading to velocity bunching in the subsequent drift space. The required amplitude of the RF electric field for maximum compression is $\sim2.2\,\rm{MV/m}$, see Fig. \ref{fig:compression}, which is achieved by driving our power efficient cavity with $51\,\rm{W}$ RF power.

To measure bunch length we use another power efficient $3\,\rm{GHz}$ RF cavity, oscillating in the $\rm{TM_{110}}$ mode, which acts as an ultrafast streak camera \cite{Thijs-UP-XVI}: the on-axis magnetic field $B_y$ deflects the electrons in the $x$-direction. The RF phase offset is chosen such that the electrons at the center of the bunch are not deflected. In this way the longitudinal bunch profile is projected as a streak on the $xy$-plane. For detection we use a micro-channel plate (MCP) with a phosphor screen that is imaged 1:1 onto a CCD camera. The relation between the bunch duration $t_b$ and the length $X_{str}$ of the streak on the phosphorous screen is obtained by integrating the Lorentz force that is acting on the electron bunch during its travel along the axis through the streak cavity. For bunch durations $t_b$ much smaller than the RF period it follows that

\begin{equation}\label{eq:tb-streak-cal}
    t_b = \mathcal{C}\frac{X_{str}}{B_0}.
\end{equation}

\begin{figure}[tbh]
    \centering
    \includegraphics[width=\columnwidth]{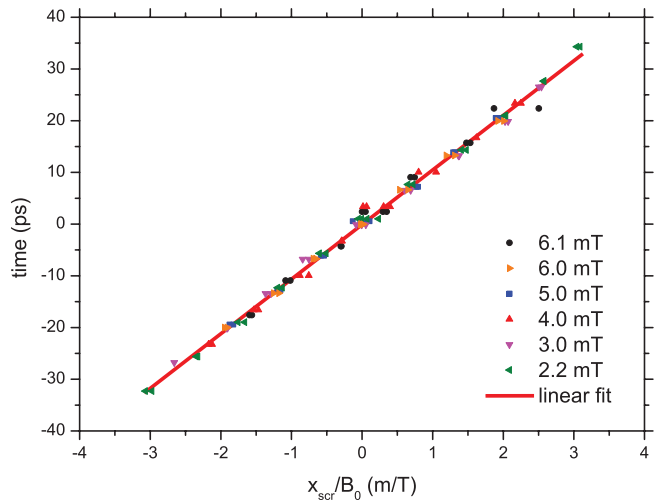}
    \caption{\it (color online) Time delay as a function of $x_{scr}/B_0$. The red line is a linear fit to the data (black dots). The thickness of the red line reflects the confidence bounds of the fit.}
    \label{fig:streak-phase}
\end{figure}

Here $B_0$ is the maximum amplitude of the magnetic field and $\mathcal{C} = \gamma m v / \left(2 \pi f_0 e L_{cav} d \right)$, with $e$ the elementary charge, $m$ the electron mass, $\gamma=\left[1-(v^2/c^2)\right]^{-1/2}$ the Lorentz factor, with $v$ the speed of the electrons and $c$ the speed of light, $d$ the distance from the exit of the streak cavity to the MCP, and $f_0 = 3\,\rm{GHz}$ the resonant frequency of the cavity. The effective cavity length is given by $L_{cav} = \int_{-\infty}^{\infty} b(z) \cos \left(\frac{\omega z}{v_z}\right) dz$, where $b(z) = B_y(z)/B_0$ is the on-axis field profile of the cavity, which is known accurately from both simulations and measurements. For our setup $\mathcal{C} = (0.90 \pm 0.01) \,10^{-11} \,\rm{s\,T\,m^{-1}}$. As an independent check we have measured the position $x_{scr}$ of the streak on the screen as a function of the RF phase offset, which is equivalent to a change in arrival time. The results for various values of $B_0$ are shown in Fig. \ref{fig:streak-phase}. The slope of the linear fit to all data yields $\mathcal{C} = (1.06 \pm 0.07)\,10^{-11}\,\rm{s\,T\,m^{-1}}$, in satisfactory agreement with the result above.

\begin{figure}[bth]
    \centering
    \includegraphics[width=\columnwidth]{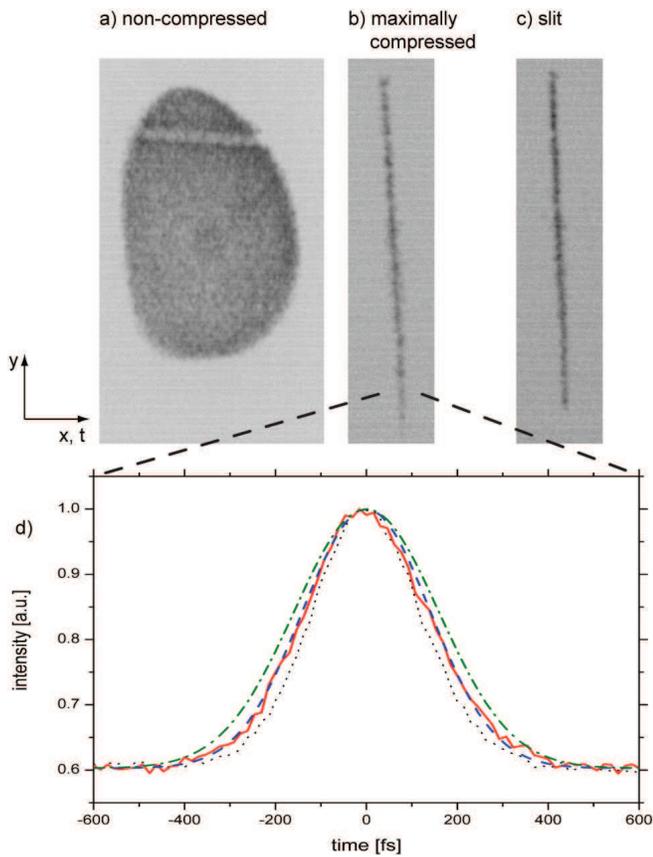}
    \caption{\it (color online.) (a) Streak of a non-compressed bunch, (b) streak of a maximally compressed bunch, (c) projection of the $10\,\rm{\mu m}$ slit when the streak cavity is off. The pronounced horizontal line in the left panel is due to a local narrowing in the slit. (d) Intensity profiles of a streak of a $67\,\rm{fs}$ bunch (red line), of the projection of the slit (black dotted line), of a Gaussian with $\sigma_{67} = \sqrt{(\sigma_{t,slit})^2 + (67\,fs)^2}$ (blue dashed line), and of a Gaussian with $\sigma_{100} = \sqrt{(\sigma_{t,slit})^2 + (100\,fs)^2}$ (green dash-dotted line).}
     \label{fig:streak}
\end{figure}

In figure \ref{fig:streak} streaks of a non-compressed and a maximally compressed $250\,\rm{fC}$ bunch are shown. Due to the streaking action the distribution on the MCP is a convolution of the transverse and longitudinal bunch profile. For Gaussian distributions the RMS width of the intensity profile on the MCP is thus given by $\sigma_{\textsc{mcp}} = \sqrt{\sigma_x^2 + X_{str}^2}$, where $\sigma_x$ is the RMS size of the bunch in the $x$-direction at the streak cavity. The spotsize of the bunch thus limits the resolution of the bunch length measurement. To increase the resolution we have placed a $10\,\rm{\mu m}$ slit in front of the streak cavity to bring down the spotsize $\sigma_x$.
\\
$\sigma_{\textsc{mcp}}$ is determined by summing five images and integrating the resulting image in the $y$-direction to increase the signal-to-noise ratio. The intensity profile thus obtained is fitted to a Gaussian, yielding $\sigma_{\textsc{mcp}}$. For streak lengths $X_{str} \approx \sigma_x$ we adopted a different procedure: single images are analyzed by taking lineouts through the streak. Each lineout is fitted to a Gaussian and shifted such that all lineouts are centered at the same position. These shifted lineouts are summed and the result is fitted to a Gaussian to obtain $\sigma_{\textsc{mcp}}$.

Figure \ref{fig:compression} shows the RMS bunch duration as a function of the RF compression field amplitude $E_0$. We are able to compress $250\,\rm{fC}$ bunches from $10\,\rm{ps}$ down to $100\,\rm{fs}$ durations. Also shown in this figure is the result of detailed particle tracking simulations with the GPT code \cite{GPT}, which take into account realistic external fields and all Coulombic interactions. Clearly the simulations agree very well with the measurements. We attribute remaining discrepancies to uncertainties in the charge, laser spotsize, and solenoid field strengths. The error in $E_0$ is mainly due to a systematic error in the detection of the absorbed RF power.
\\
At the field strength of maximal compression there is hardly any difference between the streak and the slit projection, as shown in Fig. \ref{fig:streak}. We also show in Fig. \ref{fig:streak} the streak intensity profiles of a maximally compressed bunch, the profile of the slit, and a Gaussian with $\sigma_{67} = \sqrt{(\sigma_{t,slit})^2 + (67\,\rm{fs})^2}$, where $\sigma_{t,slit}$ is the slit width converted to the time scale as if it were a streak. Because the $\sigma_{67}$-profile overlaps with the measured streak profile deconvolution of the streak yields a bunch duration $t_b = 67\,\rm{fs}$. For comparison Fig. \ref{fig:streak}d also shows a Gaussian with $\sigma_{100} = \sqrt{(\sigma_{t,slit})^2 + (100\,\rm{fs})^2}$, clearly showing that RMS bunch durations shorter well below $100\,\rm{fs}$ have been achieved.
\\
For a pump-probe UED experiment the arrival time jitter of the electron bunch with respect to the pump (generally an ultrashort laser pulse) is crucial. In our setup phase jitter of the RF compression cavity leads to changes in the average velocity of the electron bunch, resulting in arrival time jitter at the position of the sample (or streak cavity), which can be determined from the measurements shown in Fig. \ref{fig:streak-phase}. The thickness of the fitted curve reflects the confidence bounds, yielding an RMS jitter of $106\,\rm{fs}$. This arrival time jitter can be translated back into $28\,\rm{fs}$ phase jitter of the RF field in the compression cavity, which agrees with the expectation based on the $20\,\rm{fs}$ synchronization accuracy between our $3\,\rm{GHz}$ oscillator and Ti:Sapphire oscillator \cite{Fred}. The measurements in Fig. \ref{fig:streak-phase} have been performed at $E_0 = 2.9\,\rm{MV/m}$. As the arrival time jitter scales linearly with $E_0$ the RMS arrival time jitter is $80\,\rm{fs}$ at the field strength of maximum compression, i.e. $E_0 = 2.2\,\rm{MV/m}$.

To show that our bunches have sufficient charge and are of sufficient quality for single shot UED we carried out a diffraction experiment. We replaced the streak cavity by a standard calibration sample for transmission electron microscopy \cite{agar}, consisting of a $300\,\rm{\mu m}$ copper mesh, a carbon interlayer and a polycrystalline gold layer of $(9 \pm 1) \,\rm{nm}$ thickness. A third solenoid is positioned behind the sample with the MCP in its the focal plane. Figure \ref{fig:diffraction} shows a diffraction pattern, recorded with a single $200\,\rm{fC}$ electron bunch. Figure \ref{fig:diffraction} also shows the azimuthal integral of the Debye-Scherrer rings. The background due to the grid and the carbon layer has been subtracted from this curve, confirming that the rings are due to diffraction of electrons on the gold film. The curve is fitted according to kinematical diffraction theory, with the elastic scattering amplitude, the inelastic scattering amplitude, and the peak width as fit parameters. The relative positions of the Bragg peaks and their relative intensities are fixed at the theoretical values.

In conclusion we have demonstrated RF compression of non-relativistic, space-charge-dominated electron bunches to sub-$100\,\rm{fs}$ durations. Detailed charged particle simulations with the GPT code are consistent with our measurements. Our bunches are suitable for single-shot UED experiments, as we have shown by capturing a high-quality diffraction pattern from a polycrystalline gold film using a single electron bunch.\\
\newline

The authors would like to thank B.\,J. Siwick for stimulating discussions and E.\,H. Rietman, A.\,H. Kemper, and H.\,A. van Doorn for their expert technical assistance. This research is financially supported by the ``\textit{Stichting voor Fundamenteel Onderzoek der Materie'' (FOM)}, and the Dutch Technology Foundation \textit{STW}.

\end{document}